\documentclass[a4paper]{jpconf}
\usepackage{graphicx}

\usepackage{amsmath}
\graphicspath {{./figures/}}
\usepackage{caption}
\usepackage{subcaption}

\begin{document}
\title{Multicanonical simulations of the 2D spin-$1$ Baxter-Wu model in a crystal field}

\author{Nikolaos G Fytas$^1$, Alexandros Vasilopoulos$^1$, Erol Vatansever$^{2\;1}$, Anastasios Malakis$^{3\;1}$ and Martin Weigel$^{4\;1}$}

\address{$^1$ Centre for Fluid and Complex Systems, Coventry
	University, Coventry, CV1 5FB, United Kingdom}

\address{$^2$ Department of Physics, Dokuz Eyl\"{u}l University, TR-35160, Izmir, Turkey}

\address{$^3$ Department of Physics, University of Athens, Panepistimiopolis, GR 15784 Zografou, Greece}

\address{$^4$ Institut für Physik, Technische Universität Chemnitz, 09107 Chemnitz, Germany}

\ead{nikolaos.fytas@coventry.ac.uk}

\begin{abstract}
We investigate aspects of universality in the two-dimensional (2D) spin-$1$ Baxter-Wu model in a crystal field $\Delta$ using a parallel version of the multicanonical algorithm employed at constant temperature $T$. A detailed finite-size scaling analysis in the continuous regime of the $\Delta-T$ phase diagram of the model indicates that the transition belongs to the universality class of the $4$-state Potts model. The presence of first-order-like finite-size effects that become more pronounced as one approaches the pentacritical point of the model is highlighted and discussed.
\end{abstract}

\section{Introduction}
\label{sec:intro}
The spin-$1/2$ Baxter-Wu model is defined by nearest-neighbor three-spin interactions on a triangular lattice. It was first conceptualized as a model that is not invariant under inversion by Wood and Griffiths~\cite{Wood_1972} and subsequently analytically solved by Baxter and Wu who showed that its critical universality is identical to that of the $4$-state Potts model~\cite{BaxterWu1, BaxterWu2, BaxterWu3}.
The Hamiltonian of the BW model reads as
\begin{equation}\label{eq:Ham}
  \mathcal{H}
  = -J\sum_{\langle ijk \rangle}\sigma_{i}\sigma_{j}\sigma_{k},
\end{equation}
where the spins take on the values $\pm 1$, $J>0$ is the ferromagnetic exchange interaction, and the sum extends over all elementary triangles of the lattice.
Due to the fact that the triangular lattice can be decomposed into three sublattices, each spin of an elementary triangle of the lattice belongs to a different sublattice. Thus, inverting the spins of any two sublattices would result in no change in the energy. Hence, the ground state consists of one ferromagnetic state, where all spins are pointing up, and three ferrimagnetic states, where one sub-lattice is pointing up and the remaining two are pointing down.

\begin{figure}
    \centering
    \begin{subfigure}[b]{0.49\textwidth}
        \centering
    	\includegraphics[width=80mm]{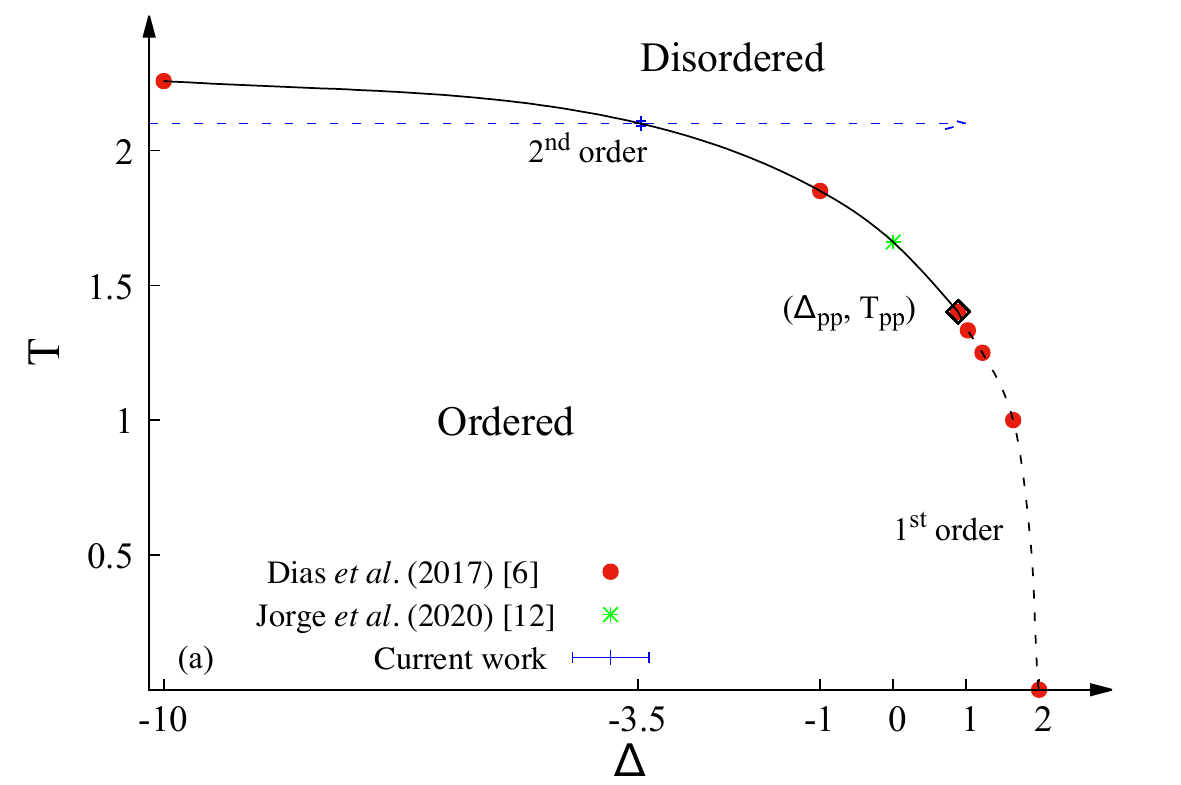}
    \end{subfigure}
    \hspace{0.01cm}
    \begin{subfigure}[b]{0.49\textwidth}
        \centering
    	\includegraphics[width=80mm]{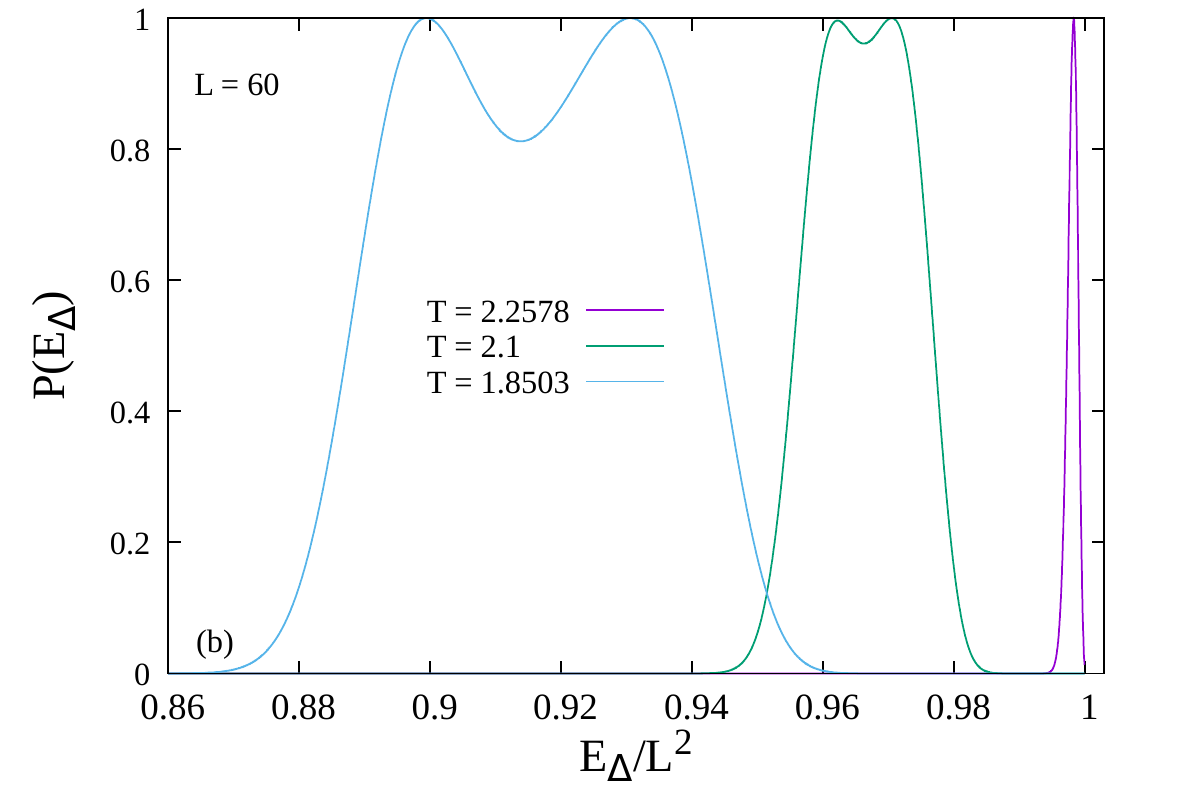}
    \end{subfigure}
    \caption{(a) Phase diagram of the 2D spin-$1$ Baxter-Wu model with the ferromagnetic and paramagnetic phases separated by a continuous transition for larger $T$ (solid line) and a first-order transition for smaller $T$ (dotted line). The line segments meet at the pentacritical point $(\Delta_{\rm pp}, T_{\rm pp}) \approx (0.8902, 1.4)$~\cite{dias17} marked by the black rhombus, where three ferrimagnetic configurations and a ferromagnetic configuration, along with that of zero spins coexist~\cite{dias17, costa04, jorge21}. The dashed arrow at $T=2.1$ indicates the temperature choice of our simulations that corresponds to $\Delta \approx -3.5$, see also figure~\ref{fig:fitD_Um}(a).
    (b) Probability density functions of the crystal-field energy $P(E_{\Delta})$ for a system of linear size $L=60$ at different temperatures in the second-order transition regime, as indicated. Note that for $T=2.2578$: $\Delta\approx -10$, and for $T=1.8503$: $\Delta\approx -1$. Finite-size effects of first-order-type (double-peak structure) appear as we lower the temperature and approach the pentacritical point.}
    \label{fig:phase-diagram-pdfs}
\end{figure}

An interesting extension of the Baxter-Wu model arises when one considers spin values $\sigma_{x} =  \{-1,0,1\}$ and includes an extra crystal field (or single-ion anisotropy) $\Delta$ coupled to $\sigma_{x}^{2}$, so that the new Hamiltonian reads as
\begin{equation}\label{eq:Ham2}
  \mathcal{H}
  = -J\sum_{\langle ijk \rangle}\sigma_{i}\sigma_{j}\sigma_{k}+\Delta\sum_{i}\sigma_{i}^{2} = E_{J}+\Delta E_{\Delta},
\end{equation}
where $E_J$ and $E_\Delta$ denote the nearest-neighbor interaction and the crystal-field energies, respectively. Note that $\Delta$ controls the density of the zero spins: when $\Delta\to -\infty$ spins are confined to the values $\pm1$, retrieving the spin-$1/2$ Baxter-Wu model.
 Correspondingly, and in full analogy to the Blume-Capel case~\cite{zierenberg17}, one expects for the model defined in equation~(\ref{eq:Ham2}) the same kind of competition between the ordered and disordered phases (mediated by the crystal field), and therefore a similar phase diagram but a different universality class is expected, see figure~\ref{fig:phase-diagram-pdfs}(a). 

Because dilution in the Potts model with four states~\cite{nienhuis79} has the same effect as $\Delta$ in the Baxter-Wu model, one should also expect the critical behavior of both models to be the same. Surprisingly though, there is an ongoing debate with respect to the degree of universality obeyed by the spin-$1$ Baxter-Wu model. This is mainly due to contradicting numerical evidence suggested over the course of last years: The results of Ref.~\cite{costa04} via renormalization group, conventional finite-size scaling, and conformal invariance techniques indicated that the critical exponents vary continuously with $\Delta$ along the second-order transition line, differently from the expected behavior of the $4$-state Potts model (with exponents $\nu=2/3$, $\alpha/\nu=1$, and $\gamma/\nu=7/4$). A similar conclusion was drawn in Ref.~\cite{costa04b}, where using importance sampling Monte Carlo simulations for the special case with $\Delta = 0$ the values $\nu = 0.617(3)$, $\alpha = 0.692(6)$, and $\gamma  = 1.13(1)$ were obtained. The complementary Monte Carlo work of Ref.~\cite{costa16} at $\Delta = -1$ and $1$ further corroborated this hypothesis. Conversely, the renormalization-group work of Dias \emph{et al.}~\cite{dias17} suggested that along the critical line, the conformal anomaly $c$ and the exponents $\nu$, $\eta$ are the same as that of the pure spin-$1/2$ Baxter-Wu model (or the $4$-state Potts model). Finally, the most recent works by Jorge \emph{et al.}~\cite{jorge20} used Wang-Landau entropic sampling simulations to probe the system at $\Delta = 0$. According to these authors the model exhibits an indeterminacy regarding the order of phase transition. 

In this conflicting situation we present here an updated finite-size scaling analysis of the model in the second-order transition regime of the phase diagram. Using multicanonical simulations outlined in section~\ref{sec:methods}, we scrutinize the critical properties of the system at the fixed temperature $T=2.1$. Via a standard finite-size scaling analysis, as discussed in section~\ref{sec:results}, we recover estimates of critical exponents that are in good agreement to those of the $4$-state Potts universality class~\cite{wu82}. Our work also features the existence of first-order-like finite-size effects in the second-order regime of the phase diagram that become more pronounced as one approaches the pentacritical point. A summary with a work-in-progress plan closes this short contribution in section~\ref{sec:summary}.

\section{Numerical approach}
\label{sec:methods}

We now turn to a description of the multicanonical (MUCA) method~\cite{berg92}. In this approach, instead of using the canonical Boltzmann weight $e^{-\beta E}$, with the inverse temperature $\beta = 1/(k_{\rm B}T)$, a correction function is introduced, designed to produce a flat histogram. For the needs of the current work, the multicanonical method was applied with respect to the crystal-field energy $E_{\Delta}$, fixing the temperature and allowing us to continuously reweight to arbitrary values of $\Delta$. To this end, the partition function 
\begin{equation}\label{Z}
	\mathcal{Z} = \sum_{\{E_J, E_{\Delta}\}} g(E_J, E_{\Delta}) e^{-\beta(E_{J}+\Delta E_{\Delta})}
\end{equation}
is generalized to
\begin{equation}\label{Z_muca}
	\mathcal{Z} _\mathrm{MUCA} = \sum_{\{E_J, E_{\Delta}\}} g(E_J, E_{\Delta}) e^{-\beta E_{J}}~W\left(E_{\Delta}\right),
\end{equation}
where $g(E_J, E_\Delta)$ is the two-parametric density of states.

It follows that 
\begin{equation}\label{P_muca}
	P_\mathrm{MUCA}(E_J, E_\Delta)=\frac{g(E_J, E_{\Delta}) e^{-\beta E_{J}}W(E_{\Delta})}{\mathcal{Z} _\mathrm{MUCA}},
\end{equation}
where $P_\mathrm{MUCA}$ is the equilibrium probability distribution. In order to produce a flat $E_{\Delta}$ histogram, by carrying out a summation with respect to $E_J$, the modified weight should be given by
\begin{equation}\label{W_muca}
	W(E_{\Delta}) \propto \mathcal{Z} _\mathrm{MUCA} \left[ \sum_{E_J} g(E_J, E_\Delta) e^{-\beta E_J}  \right]^{-1}.
\end{equation}

These weights can be calculated in an iterative fashion starting with an initial guess.
At the $n^\text{th}$ step spins are flipped using the weights $e^{-\beta E_{J}}W^{(n)}\left(E_\Delta\right)$ and the histogram $H^{(n)}(E_\Delta)$ of the energies $E_\Delta$ is sampled. After a specified number of spin-flip attempts the histogram is used to recalibrate the weights via $W^{(n+1)}\left(E_\Delta\right) = W^{(n)}\left(E_\Delta\right)/H^{(n)}(E_\Delta)$.
The process is completed when a flat-enough histogram has been sampled, after which a series of production runs is carried out. At each step the normalized histogram $H^{(n)}_{\mathrm{norm}}(E_\Delta)$ satisfies the equation
\begin{equation}\label{H_norm}
\langle H^{(n)}_{\mathrm{norm}}(E_\Delta)\rangle = P^{(n)}(E_\Delta) = \frac{1}{\mathcal{Z}_\mathrm{MUCA}}\sum_{E_J} g(E_J, E_{\Delta})e^{-\beta E_{J}} W^{(n)}(E_\Delta) \propto \frac{W^{(n)}(E_\Delta)}{W(E_\Delta)},
\end{equation}
justifying the scheme for updating the weights using sampled histograms.

We employed a parallel implementation of the multicanonical method~\cite{zierenberg13, gross18}, guided by its already successful application in the study of the Blume-Capel model~\cite{zierenberg17,zierenberg15,fytas18}. In this setup weights are distributed to parallel workers, each one producing a histogram. At the end all histograms are added and the resulting total histogram is then used to recalibrate the weights. Our simulations were implemented on an ``Nvidia Tesla K80'' Graphics Processing Unit boosting the numerical capacity to $26\; 624$ simulations of uncorrelated systems in parallel. Finally, the  histogram's flatness was tested using the Kullback-Leibler divergence as discussed in reference~\cite{gross18}. 

The numerical protocol described above was applied on triangular lattices with periodic boundary conditions. As the system has in addition to the ferromagnetic phase three different ferrimagnetic phases, the allowed values of the linear size of the lattice $L$ must be a multiple of three. In this way, all ground states of the infinite system would fit on any finite lattice~\cite{costa16}. In the course of our simulations we considered linear sizes within the range $L = \{12 - 72\}$ respecting this constraint. The bulk of our simulations was performed at $T=2.1$. Other values of $T$ closer to the pentacritical point are currently being considered and will be reported elsewhere~\cite{vasilopoulos21}. 

In the framework of the multicanonical approach it is natural to compute $\Delta$-derivatives of observables rather than the usual $T$-ones. For instance, in place of the usual specific heat one may define a specific-heat-like quantity~\cite{zierenberg15}
\begin{equation}
	C_\Delta = \frac{1}{N} \frac{\partial E_J}{\partial\Delta}
	=
	-\left[ \left\langle E_J E_\Delta \right\rangle - \left\langle E_J \right\rangle \left\langle E_\Delta \right\rangle \right]/(NT),
\end{equation}
\noindent where $N = L^2$, the number of lattice sites. Other finite-size scaling observables include the partial derivative of the logarithm of the $n^{\rm th}$ power of the magnetization $m$ ($\partial\ln{\langle m^n \rangle}/\partial \Delta$), the magnetic susceptibility ($\chi$), as well as the fourth-order Binder cumulant of the magnetization ($U_{\rm m}$)~\cite{jorge20,zierenberg15,ferrenberg_landau91}.

\begin{figure}
    \centering
    \begin{subfigure}[b]{0.49\textwidth}
        \centering
    	\includegraphics[width=80mm]{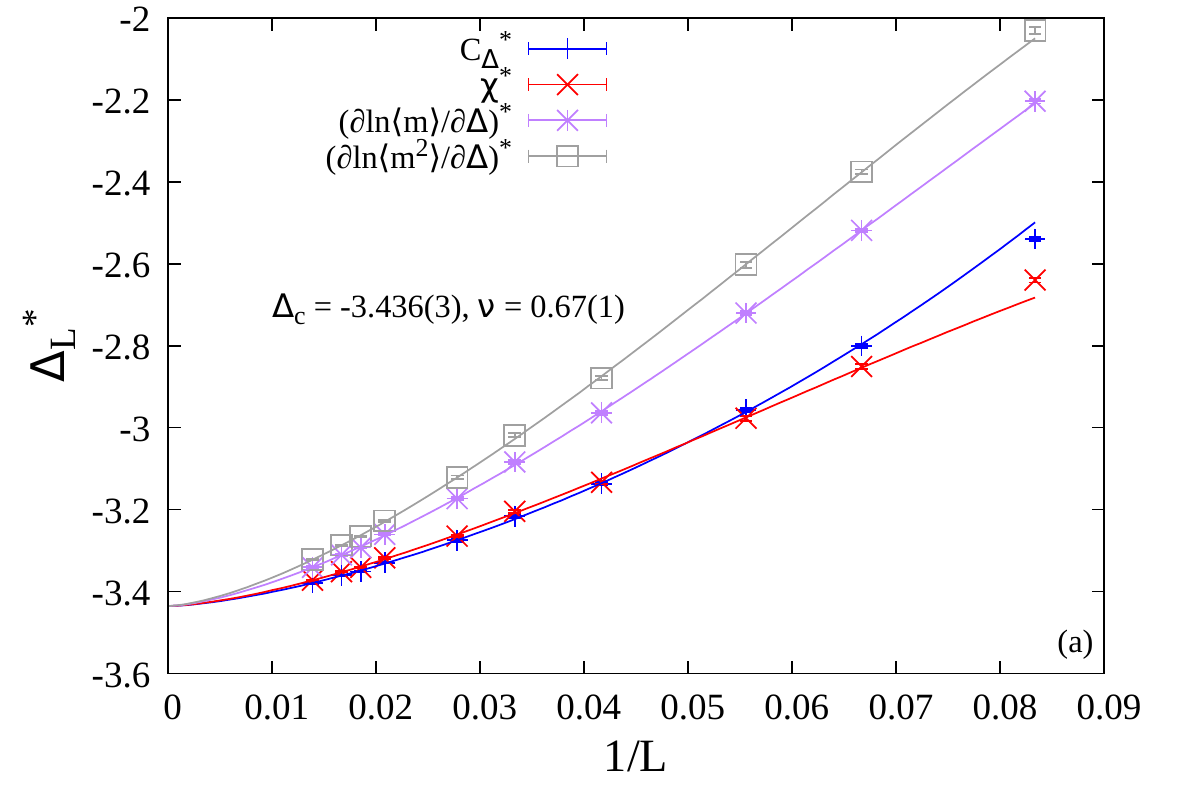}
    \end{subfigure}
    \begin{subfigure}[b]{0.49\textwidth}
        \centering
    	\includegraphics[width=80mm]{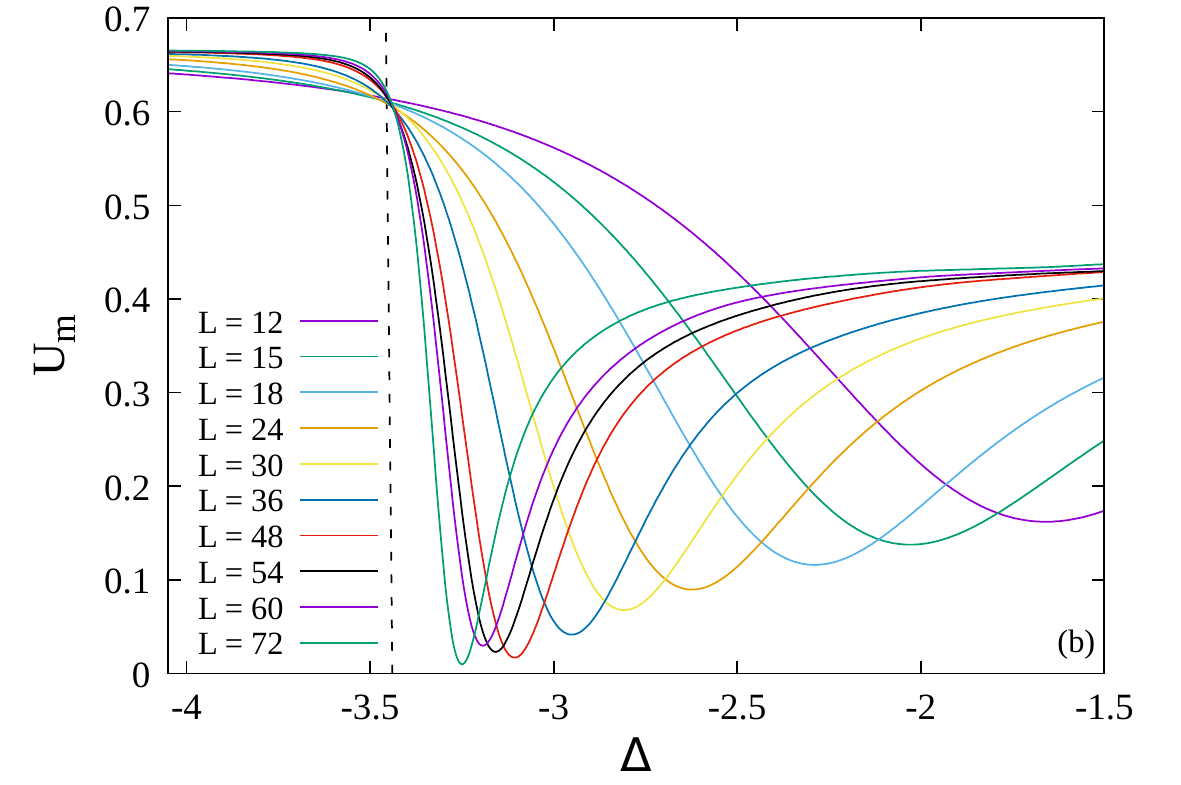}
    \end{subfigure}
    \caption{(a) Shift behaviour of several pseudocritical fields $\Delta^{\ast}_{L}$ as indicated in the legend. (b) Crossings of the fourth-order magnetization's Binder cumulant for different system sizes.}  
    \label{fig:fitD_Um}
\end{figure}

\section{Results}
\label{sec:results}

We start with some typical illustrations of the (reweighted) probability density function $P(E_{\Delta})$ (normalized to unity). It is well known that a single-peak structure in $P(E_{\Delta})$ is characteristic of a continuous transition, whereas a double-peak structure signals the emergence of a first-order phase transition~\cite{zierenberg15,binder84,binder87}.  Figure~\ref{fig:phase-diagram-pdfs}(b) depicts $P(E_{\Delta})$ for a system with $L = 60$ at several temperatures (see caption). The distributions were reweighted close to their respective critical crystal field according to the calculations of Dias \emph{et al.}~\cite{dias17} and the current work. Some comments are in order: (i) For large-enough $T$ a single peak is observed, but as we lower $T$ (and increase $\Delta$ up to $-1$) first-order-like characteristics appear similar to the observations by Jorge \emph{et al.}~\cite{jorge20} at $\Delta = 0$. (ii) These effects become more pronounced as we approach the pentacritical point and lead to corrections in finite-size scaling in this vicinity. Clearly, a refined investigation of the surface tension and latent heat of the transition is called for, as it is quite often the case that pseudo-first-order signatures emerge at moderate system sizes in many spin models with complex interactions, see~\cite{fytas_rfim} and references therein. Some first analysis along these lines indicates that these are indeed mere finite-size effects that disappear in the thermodynamic limit~\cite{vasilopoulos21}. 

We proceed now with a finite-size scaling analysis at $T=2.1$ in the second-order regime. For the fitting procedure discussed below we restricted
ourselves to data with $L\geq L_{\rm min}$. As usual, to determine an
acceptable $L_{\rm min}$ we employed the standard $\chi^2$ test for goodness of the fit. Specifically, we considered a fit as being fair only if $10\% < Q < 90\%$, where $Q$ denotes the probability of finding a $\chi^2$ value which is even larger than the one actually found from our data~\cite{press92}.

To extract the critical crystal field $\Delta_{\rm c}$ and a first estimate of the correlation-length exponent $\nu$ we study the shift behaviour of several suitable pseudocritical fields, $\Delta_{L}^{\ast}$, deduced from the peak locations of the relevant observables $C_{\Delta}$, $\chi$, and $\partial\ln{\langle m^n \rangle}/\partial \Delta$. Figure~\ref{fig:fitD_Um}(a) features the joint fit~\cite{dias17,ferrenberg_landau91}
\begin{equation}
\label{eq:shift}
	\Delta^{\ast}_{L} = \Delta_{\rm c} + bL^{-1/\nu}(1+b'L^{-\omega}),
\end{equation}
over four different data sets and with $L_{\rm min} = 15$. Note that $b$ and $b'$ are non-universal coefficients and the corrections-to-scaling exponent was set to the well-known value $\omega=2$~\cite{dias17,costa16,alcaraz97,alcaraz99}. We obtain the estimates $\Delta_{\rm c}=-3.436(3)$ and $\nu = 0.67(1)$, the latter being in very good agreement to the value $\nu = 2/3$ of the $4$-state Potts model~\cite{BaxterWu1,BaxterWu2}. Typical curves of the Binder cumulant $U_{m}$ are shown in figure~\ref{fig:fitD_Um}(b) where the crossing point agrees (within some minor finite-size effects) with the value $\Delta = -3.436$, as marked by the vertical dashed line.

\begin{figure}
    \centering
    \begin{subfigure}[b]{0.49\textwidth}
        \centering
    	\includegraphics[width=80mm]{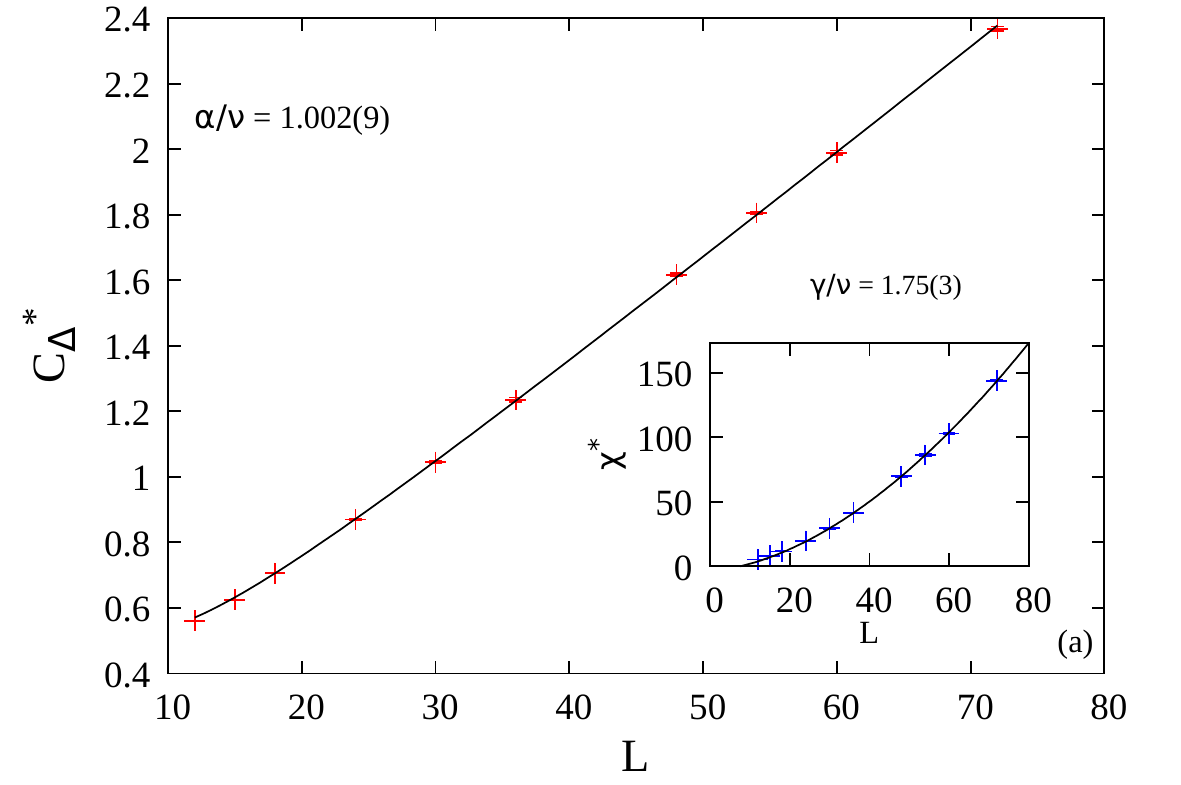}
    \end{subfigure}
    \begin{subfigure}[b]{0.49\textwidth}
        \centering
    	\includegraphics[width=80mm]{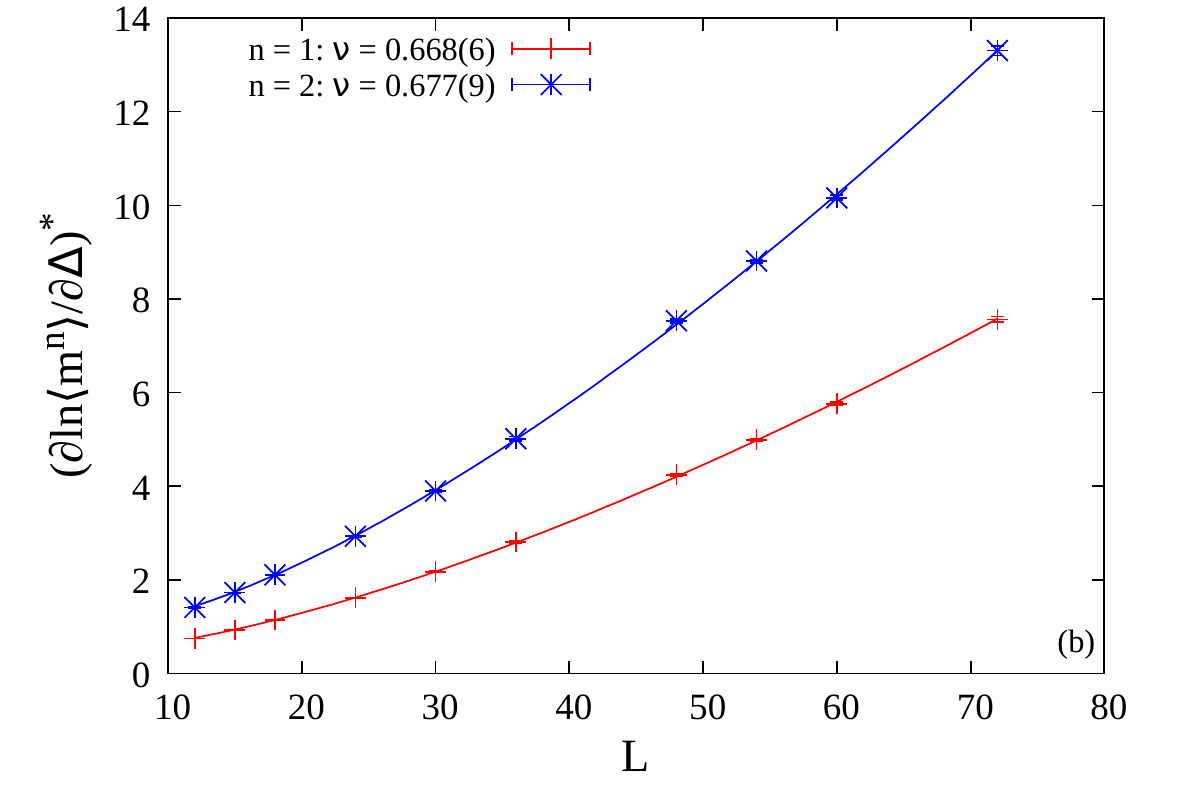}
    \end{subfigure}
    \caption{Finite-size scaling behaviour of $C_\Delta^\ast$ (main panel) and $\chi^\ast$ (inset) in panel (a) and of $(\partial\ln{\langle m^n \rangle}/\partial \Delta)^{\ast}$ in panel (b).}
    \label{fig:scaling_c-x-dlnm}
\end{figure}

Further, in figure~\ref{fig:scaling_c-x-dlnm}(a) we present the finite-size scaling behaviour of the maxima of the specific-heat-like quantity (main panel) and the  magnetic susceptibility (inset). The solid lines are fits of the form~\cite{zierenberg17}
\begin{equation}
\label{eq:spec-heat}
	 C_\Delta^{\ast} \sim L^{\alpha/\nu}(1+b'L^{-\omega}) \;\; ; \;\; \chi^{\ast} \sim L^{\gamma/\nu}(1+b'L^{-\omega}),
\end{equation}
using $L_{\rm min} = 18$ and $30$ respectively. The resulting estimates $\alpha/\nu = 1.002(9)$ and $\gamma/\nu = 1.75(3)$ are clearly compatible to the exact values $1$ and $7/4$ of the $4$-state Potts model~\cite{BaxterWu1,BaxterWu2}. Finally, in figure~\ref{fig:scaling_c-x-dlnm}(b) we provide an additional verification of the critical exponent $\nu$ via the maxima of the logarithmic derivatives of the order parameter. A fit of the form~\cite{ferrenberg_landau91} 
\begin{equation}
\label{eq:lnm}
	 (\partial\ln{\langle m^n \rangle}/\partial \Delta)^{\ast} \sim L^{1/\nu} (1+b'L^{-\omega}),
\end{equation}
where $L_{\rm min} = 18$ gives $\nu = 0.668(6)$ and $0.677(9)$ for $n=1$ and $2$ respectively, in further support of the $4$-state Potts universality class.

\section{Summary and outlook}
\label{sec:summary}

Using parallel multicanonical simulations and finite-size scaling analysis we studied the universality aspects of the 2D spin-1 Baxter-Wu model at the second-order transition regime of the $\Delta-T$ phase boundary. Our estimates for the critical exponents provide evidence that this ferromagnetic-paramagnetic transition belongs to the universality class of the 4-state Potts model. In order to capture possible finite-size effects of first-order type as already discussed recently~\cite{jorge20}, we monitored the structure of the crystal-field energy probability density function. Indeed, on lowering the temperature a double-peak structure in the energy probability density function is observed, which becomes more pronounced as the pentacritical point is approached. The actual role of these finite-size effects including simulation of larger system sizes and a systematic scaling analysis of the respective surface tension and latent heat of this pseudo-first-order transition are currently under study by our group~\cite{vasilopoulos21}. Some future plans include a dedicated study of the location and universality principle of the pentacritical point, as well as a high-accuracy reproduction of the model's phase diagram.

\ack

N G Fytas would like to thank J A Plascak for useful communication on the topic. We acknowledge the provision of computing time on the parallel computer cluster \emph{Zeus} of Coventry University.

\section*{References}

\bibliography{iopart-num}

\providecommand{\newblock}{}
\begin{thebibliography}{10}
\expandafter\ifx\csname url\endcsname\relax
  \def\url#1{{\tt #1}}\fi
\expandafter\ifx\csname urlprefix\endcsname\relax\def\urlprefix{URL }\fi
\providecommand{\eprint}[2][]{\url{#2}}

\bibitem{Wood_1972}
Wood D~W and Griffiths H~P 1972 {\em Journal of Physics C: Solid State
  Physics\/} {\bf 5} L253

\bibitem{BaxterWu1}
Baxter R and Wu F 1973 {\em Phys. Rev. Lett.\/} {\bf 31} 1294

\bibitem{BaxterWu2}
Baxter R 1974 {\em Aust. J. Phys.\/} {\bf 27} 357

\bibitem{BaxterWu3}
Baxter R and Wu F 1974 {\em Aust. J. Phys.\/} {\bf 27} 369

\bibitem{zierenberg17}
Zierenberg J, Fytas N~G, Weigel M, Janke W and Malakis A 2017 {\em Eur. Phys. J
  Special Topics\/} {\bf 226} 789–804

\bibitem{dias17}
Dias D~A, Xavier J~C and Plascak J~A 2017 {\em Phys. Rev. E\/} {\bf 95} 012103

\bibitem{costa04}
Costa M~L~M, Xavier J~C and Plascak J~A 2004 {\em Phys. Rev. B\/} {\bf 69}
  104103

\bibitem{jorge21}
Jorge L, Martins P, DaSilva C, Ferreira L and Caparica A 2021 {\em Physica A\/}
  {\bf 576} 126071

\bibitem{nienhuis79}
Nienhuis B, Berker A~N, Riedel E~K and Schick M 1979 {\em Phys. Rev. Lett.\/}
  {\bf 43}(11) 737--740

\bibitem{costa04b}
Costa M~L~M and Plascak J~A 2004 {\em Braz. J. Phys.\/} {\bf 34} 419–421

\bibitem{costa16}
Costa M~L~M and Plascak J~A 2016 {\em J. Phys.: Conf. Ser.\/} {\bf 686} 012011

\bibitem{jorge20}
Jorge L, Ferreira L and Caparica A 2020 {\em Physica A\/} {\bf 542} 123417

\bibitem{wu82}
Wu F~Y 1982 {\em Rev. Mod. Phys.\/} {\bf 54} 235

\bibitem{berg92}
Berg B~A and Neuhaus T 1992 {\em Phys. Rev. Lett.\/} {\bf 68} 9–12

\bibitem{zierenberg13}
Zierenberg J, Marenz M and Janke W 2013 {\em Comput. Phys. Commun.\/} {\bf 184}
  1155–1160

\bibitem{gross18}
Gross J, Zierenberg J, Weigel M and Janke W 2018 {\em Computer Physics
  Communications\/} {\bf 224} 387–395

\bibitem{zierenberg15}
Zierenberg J, Fytas N~G and Janke W 2015 {\em Phys. Rev. E\/} {\bf 91} 032126

\bibitem{fytas18}
Fytas N~G, Zierenberg J, Theodorakis P~E, Weigel M, Janke W and Malakis A 2018
  {\em Phys. Rev. E\/} {\bf 97} 040102

\bibitem{vasilopoulos21}
Vasilopoulos A, Fytas N~G, Vatansever E, Malakis A and Weigel M 2021 {\em (in
  preparation)\/}

\bibitem{ferrenberg_landau91}
Ferrenberg A~M and Landau D~P 1991 {\em Phys. Rev. B\/} {\bf 44}(10) 5081--5091

\bibitem{binder84}
Binder K and Landau D~P 1984 {\em Phys. Rev. B\/} {\bf 30} 1477

\bibitem{binder87}
Binder K 1987 {\em Rep. Prog. Phys.\/} {\bf 50} 783

\bibitem{fytas_rfim}
Fytas N~G, Malakis A and Eftaxias K 2008 {\em J. Stat. Mech.\/}  P03015

\bibitem{press92}
Press W, Teukolsky S, Vetterling W and Flannery B 1992 {\em Numerical Recipes
  in C\/} 2nd ed 27 (Cambridge: Cambridge University Press)

\bibitem{alcaraz97}
Alcaraz F~C and Xavier J~C 1997 {\em J. Phys. A: Math. Gen.\/} {\bf 30} L203

\bibitem{alcaraz99}
Alcaraz F~C and Xavier J~C 1999 {\em J. Phys. A: Math. Gen.\/} {\bf 32} 2041

\end{thebibliography}

\end{document}